\documentclass[fleqn,usenatbib]{mnras}

\usepackage{newtxtext,newtxmath}

\usepackage[T1]{fontenc}

\DeclareRobustCommand{\VAN}[3]{#2}
\let\VANthebibliography\thebibliography
\def\thebibliography{\DeclareRobustCommand{\VAN}[3]{##3}\VANthebibliography}

\usepackage{graphicx}	
\usepackage{amsmath}	
\usepackage{dsfont}
\usepackage{graphicx}
\newcommand{\given}{\, | \,}
\newcommand{\homo}{{\em Homo}}
\newcommand{\sapiens}{{\em Homo sapiens}}

\title[Bayesian anthropic inference]{On the foundations of anthropic reasoning:  Implications for life across the universe}

\author[B. J. Brewer et al.]{
Brendon J. Brewer,$^{1}$
Geraint F. Lewis,$^{2}$\thanks{E-mail: geraint.lewis@sydney.edu.au}
Jesse Goodman$^{1}$
\\
$^{1}$Department of Statistics, The University of Auckland, Private Bag
92019, Auckland, 1142, New Zealand\\
$^{2}$Sydney Institute for Astronomy, School of Physics, A28, The
University of Sydney, NSW, 2006, Australia
}

\date{To be submitted} 

\pubyear{\the\year{}}

\begin{document}
\label{firstpage}
\pagerange{\pageref{firstpage}--\pageref{lastpage}}
\maketitle

\begin{abstract}
With only one confirmed example, namely the Earth, the abundance of life across the universe remains deeply uncertain, and local evidence about Earth’s particular circumstances can significantly shape our knowledge of life elsewhere. Such extrapolation is difficult not only because of this sample size of one, but also due to observer selection effects discussed under the Anthropic Principle. In this paper, we describe and apply a framework called Full Indexical Conditioning (FIC) for performing Bayesian inference in this situation. Inspired by, but distinct from, Full Non-Indexical Conditioning, FIC conditions on all evidence available to the reasoner while considering that this evidence could have been otherwise. We address the reference class problem associated with such reasoning, demonstrating that substantive conclusions are unaffected by the choice of reference class provided it is sufficiently large. We apply FIC to several problems in astrobiology, including the Doomsday argument, the early abiogenesis time, and our location around a G-type rather than M-type star, clarifying the foundations of these arguments. We conclude that the early abiogenesis time does not rule out pessimistic possibilities for life’s abundance, and that while life around M stars may be uncommon, its temporal distribution could extend far into the future. Ultimately, FIC offers a clear conceptual framework, grounded in straightforward probability theory, for tackling problems involving anthropic selection effects.
\end{abstract}




\section{Introduction}\label{sec:introduction}
Few questions in astronomy are as fundamental or potentially impactful as whether life exists beyond the Earth, and how common it may be across the universe. Yet any attempt to answer this question must confront a peculiarity of our evidence, namely that we can only observe the universe from a vantage point in which life, in particular intelligent life, has already emerged. This selection effect, that our existence is not a random observation of the cosmos, is at the heart of what is known as Anthropic Reasoning \citep{carter1983anthropic}.  This constrains the kinds of universes, environments, and evolutionary histories we are capable of observing. The challenge, therefore, is to determine what information about life elsewhere can legitimately be inferred from the fact that life exists here
and the circumstances of its development.

This challenge is compounded by what is known as the reference class problem, which shows up in many attempts to formalise
anthropic reasoning \citep{bostrom2002anthropic}.
To draw probabilistic inferences from our existence, we must first specify the population with respect to which we regard ourselves as a typical observer. 
Should we consider ourselves as representative of all observers that could exist, from all observers that currently do exist, from all intelligent observers, or from some other more narrowly defined class? 
Different choices of reference class can yield substantially different conclusions about the likelihood of life elsewhere, and hence there is no clear picture of what we have learnt. 
The problem is therefore not simply one of accounting for observational selection, but of determining the appropriate population over which that selection should be applied.

In this paper we present a unified framework called Full Indexical Conditioning (FIC) to address the various challenges outlined above.
FIC is related to, but distinct from,
Full Non-Indexical Conditioning (FNC) as
proposed by \citet{neal2006puzzles}.
The core idea is that inference is conditioned on all evidence that is available to a reasoner (including, in principle, all of their memories), and incorporating the possibility that this evidence could have been some other set of evidence; that is, the inferential procedure should be able
to handle our specific evidence as well as the evidence
of other humans and hypothetical alien beings.
The layout of this paper is as follows:
In Section~\ref{sec:framework} we present the probabilistic framework of FIC, whilst Section~\ref{sec:examples} presents a series of examples, including the Doomsday argument, to illustrate the application of FIC. Section~\ref{sec:astrobiology} considers the application of FIC to astrobiological situations, and we conclude in Section~\ref{sec:conclusion}.

\section{Probability Framework}\label{sec:framework}

\subsection{Bayesian Inference}\label{sec:bayes}
Throughout this paper we use Bayesian inference, which models
uncertainties using probability theory
\citep{jaynes2003probability, o2004kendall}. Typically, the
initial uncertainty about unknown quantities $\theta$ is described by
a prior distribution $p(\theta)$ over the set of possible
values being considered. A `sampling distribution' $p(D \given \theta)$
is used to describe prior knowledge
about what datasets would be
expected under different values of $\theta$ \citep{caticha2006updating}.
When the data $D$ becomes known, the uncertainty about $\theta$ is
described by the {\em posterior distribution}, proportional
to the prior multiplied by the likelihood function:
\begin{align}
p(\theta \given D) &\propto p(\theta) p(D \given \theta).
\end{align}
The likelihood function is given by the sampling distribution
evaluated at the observed value of $D$, and describes how
the probability of the specific data $D$ varies as a function
of the hypothesised value of $\theta$.
The purpose of FIC is to derive the correct form of the likelihood
function for problems involving anthropic selection effects.

In this paper all analyses are presented at this parameter
estimation level and we do not consider any calculations
that fall under the banner of model comparison.

\subsection{Chain-Product Rule and Bernoulli Trials}
In this paper we make repeated use of two results
from probability theory, which we outline here.
The first we call the chain-product rule, a special
case of the product rule of probability theory. The product rule usually
reads
\begin{align}
P(A, B \given C) &= P(A \given C) P(B \given A, C),
\end{align}
and gives the probability of both $A$ and $B$ in terms of the
probability of $A$ alone and the conditional probability of $B$ given $A$.
All of this occurs within the context of $C$, which appears in the right
hand side of every term.
This equation
applies both to epistemological/Bayesian probabilities describing
the plausibility of propositions, and also to relative frequencies \citep[when
a problem contains them;][]{knuth2012foundations}.
A special case involves nested propositions or sets $A \subseteq B \subseteq C$, which we call {\em chains}, following \citet{knuth2012foundations}.
In this case, the product rule formula simplifies to
\begin{align}
P(A \given C) &= P(B \given C)P(A \given B).
\end{align}
As an example of this equation applied to
relative frequencies, consider the subset of the world's population born in Sydney,
the subset born in Australia, and the set of all people.
By the product rule, we can decompose the proportion of all people
born in Sydney into the proportion born in Australia, multiplied
by the proportion of Australians born in Sydney:
\begin{align}
f({\rm Sydney} \given {\rm World}) &= f({\rm Aus} \given {\rm World})f({\rm Sydney} \given {\rm Aus}).
\end{align}
Here we have used the symbol $f$ for relative frequencies (a convention
we adopt throughout), whereas $P$ will be reserved for epistemological
probabilities and $p$ for probability distributions henceforth.
In astrobiology, the most famous example of the chain-product rule,
applied to relative frequencies, is the Drake equation \citep{1965cae..book..323D}, which involves a large
chain of subsets, such as the proportion of planets
that host life and the proportion of such planets
that eventually release detectable signs
of their existence into space.

A particular result about `Bernoulli trials'
(experiments which can result in a zero or a one,
with a certain probability)
is also used repeatedly throughout this paper.
If an event has probability $\epsilon$ of successfully occurring at each
independent
attempt and $N$ attempts are made, the probability of
one or more successes is given by one minus the probability
that all attempts fail:
\begin{align}
P(\textnormal{at least one success}) &= 1 - (1 - \epsilon)^N.
\end{align}
If the expected number of successes
$N\epsilon$ is very small we can use the Taylor expansion
around $\epsilon=0$ to say
\begin{align}
P(\textnormal{at least one success}) &\approx N\epsilon.
\end{align}
That is, the probability of a success scales linearly
with the number
of attempts.
In most examples in this paper
$N\epsilon$ is small enough that this should scarcely be considered an approximation at all.

\subsection{Anthropic Principles}
A significant difficulty in reasoning about the abundance
of life in the universe is the fact that the Earth and its lifeforms
are the only known example, and that we are only able to
perform any inference by virtue of our existence.
Many authors \citep[e.g.][]{carter1983anthropic} have argued
that our own existence should tell us very little about the
abundance of life in the universe.

Several formal principles have been proposed for handling evidence
involving anthropic selection effects in terms of Bayesian probability theory \citep{bostrom2002anthropic}.
We briefly introduce these principles here.
Throughout this paper, we make use of a hypothetical reasoner
named Alice, who wishes to make inferences about the universe
from the data of her own memories and circumstances.

The Self-Indication Assumption (SIA) asserts that hypotheses that make
Alice's particular existence more likely are supported by the data
of her existence. SIA is consistent with all statistical methods that
are based on likelihood functions (Bayesian or otherwise).
However, SIA has a tendency to increase the
estimated population of the universe, leading to the
`Presumptuous Philosopher' problem, where reasoners who
use SIA on its own strongly prefer theories where the
universe is highly populated because it creates more
chances for their specific existence to arise.
Conversely, the Self-Sampling Assumption (SSA) \citep{bostrom2002anthropic}
states that if Alice wants to do inference
about some question, she
should consider herself
as having been `randomly selected' from some greater population of
observers. This often creates a tendency pulling in the opposite direction
to SIA, often introducing likelihood
factors of the form $1/N$ where $N$ is the size
of some population. However, a difficulty with SSA is the so called
reference class problem, relating to the choice of the boundary
of the population from which Alice has been sampled.

In 2006, \citet{neal2006puzzles} introduced `Full Non-Indexical Conditioning'
(FNC) as an alternative to both SIA and SSA.
FNC treats Alice's data as a global existence proposition of the
form `Alice's memories exist somewhere' but does not
consider indexical information --- that Alice is the one who
has those memories.
The philosophy behind FNC is straightforward and compelling --- that one
should condition on all available data. In principle this includes
every fact Alice knows, including her own biographical details.
If some data
(e.g., as Neal suggests, the name of Alice's kindergarten teacher)
is irrelevant, it should not affect the conclusions.
FNC removes issues of reference classes entirely, and often
gives results equivalent to applying SIA and SSA together.
However, as we will show in Section~\ref{sec:abundance},
FNC gives (in our view) inappropriate conclusions on astrobiological
problems, resembling the Presumptuous Philosopher which appears under
SIA.

In his paper, \citet{neal2006puzzles} briefly mentions a possible alternative to
FNC which he termed Full Indexical Conditioning (FIC).
This augments the FNC global existence proposition with
another SSA-like term, so the total data becomes
$({\rm Alice~exists~AND} {\rm~Alice~selected})$.
Throughout this paper we abbreviate this proposition to
$({\rm AE}, {\rm AS})$. The probability
that Alice exists and is selected may be factorised by the
product rule, giving
\begin{align}
P({\rm AE}, {\rm AS})
    &= P({\rm AE})P({\rm AS} \given {\rm AE}) \\
    &= (\textnormal{FNC term})\times(\textnormal{SSA term}).
\end{align}
By including the sampling step, FIC removes the
tendency of FNC to inflate the population of the universe
(as we demonstrate in Section~\ref{sec:abundance}).
This decomposition also shows that, if one accepts FIC, one also accepts
SSA, and can use it, understanding that it pre-conditions on
the reasoner's existence which thus should already be taken into account 
when assigning any prior distributions.

The inclusion of the SSA-style factor within FIC reintroduces
concerns about the reference class from which Alice
imagines herself as having been selected.
We discuss reference classes
in detail in the context of our first example, presented
in Section~\ref{sec:countries}, concluding that
reference classes are a matter of (i) the choice of coordinate
system on the space of the unknown parameters, and (ii)
the factorisation of unknown proportions into the products
of other unknown proportions. We ultimately conclude that reference class
choices do not affect substantive conclusions, provided
the reference class is large enough for the question being
asked. In this paper we present intuitive arguments for this
conclusion, leaving a more formal development of the theory of reference
classes to a future contribution.

Note that the sampling metaphor in SSA and FIC does not require a literal
random allocation process to take place. It is simply a language
convention surrounding the use of uniform
probabilities, and can also be justified
by epistemic symmetry considerations \citep{marquis1825essai,10.1093/mnras/staf1398}.
For example, in the Monty Hall problem \citep{engel1991monty}, it is valid to assign
initial probabilities $(1/3, 1/3, 1/3)$ for the prize being behind
each of the three doors without believing that some `random mechanism'
was used to decide where the prize was placed.
A minimal interpretation of the self sampling metaphor
in SSA and FIC is that `before' knowing the contents of your
memories, you do not know what they will be,
but they must be one out of the set of available
memory states produced by the universe.
However, the idea of being in a prior state of knowledge
where you do not know your own memories is
somewhat strange. Instead,
one can imagine that a professor of computational
philosophy is writing a computer program capable
of analysing the user's memories and drawing
conclusions from them. The sampling step implies that
if the unknown parameters
would lead to $N$ `observers'
in the universe, then the professor would not
be able to predict which of the $N$ observers is using
the program.

Under this view, reference classes do not necessarily
have any intrinsic connection to consciousness.
It is possible to analyse the data about a particular rock,
and use the data to make conclusions about rocks in general.
The reference class
is not connected to whether or not rocks are conscious,
but instead refers to which set of objects in the universe
a particular rock can be considered representative of.
Questions about what types of beings are conscious
are merely one possible {\em application} of this
type of reasoning.
Fortunately, we can reason about life in the universe without having
to know whether LLMs, dogs, ants, or possible
aliens are conscious.
However, if the possible consciousness of
different types of beings is of interest, and FIC is applied
as in Section~\ref{sec:countries},
the fact that you are a conscious human is weak evidence
that humans are a large proportion of the conscious population,
with solipsism being the maximum likelihood estimate.

\section{Examples}\label{sec:examples}

\subsection{Populations of Countries}\label{sec:countries}
We now demonstrate FIC on an illustrative example about the
populations of various countries.
Suppose the population of Wales is $N_{\rm Wales}$ and the
population of the USA is $N_{\rm USA}$. Alice
was born in Wales\footnote{If she had been, her name might be spelt
Alis or Alys, but we will stick to the more common spelling for familiarity.}, and wishes to infer something about the
population sizes of these two countries. Of course, there is much more data available
than just having been born in Wales, which dominates the inference
in real life. However, the likelihood term from having been born in
Wales is still present in principle.
In an astrobiology context, there
is not a lot of external information, and the local information of the
reasoner's circumstances can be relatively significant, so it is important
to clarify how it works.
If it is difficult to imagine what the evidence of Alice's Welsh
birth tells her, it can be helpful to mentally remove
the external
information that we have about country populations.
For instance, you can imagine that
Alice has lived in a jail cell her entire life and knows about the existence
of people and countries from books, but never happened to learn the populations
of the countries.

The FIC likelihood is given by the joint probability of Alice's
overall existence 
{\em and} having been `selected' from the available
population, similar to SSA.
To define Alice's existence, all biographical details
and memories available to her can be included, in principle \citep{neal2006puzzles}. In writing down the probability
that Alice exists, we imagine that each of the $N_{\rm Wales}$
Welsh people has a very small chance $\epsilon$ of being
precisely
Alice in the sense of having the same middle name, birthday,
eye colour, high school science teacher, childhood pet's name,
etc.
Thus, the likelihood function is
\begin{align}
P({\rm AE, AS})
    &= P({\rm AE})P({\rm AS} \given {\rm AE}) \\
    &= \left(N_{\rm Wales}\epsilon\right)\left(\frac{1}{N_{\rm Wales} + N_{\rm USA}}\right)\\
    &\propto \frac{N_{\rm Wales}}{N_{\rm Wales} + N_{\rm USA}}.\label{eqn:wal_likelihood1}
\end{align}
To save space in our notation,
we exclude the conditioning on parameters
from the notation when writing down
$P({\rm AE, AS})$. The fact that
unknown parameters appear in the result is enough to
know that this probability is conditional on those parameters.

The result of Equation~\ref{eqn:wal_likelihood1}
is traditionally associated with SSA-style
arguments; however, only the denominator is related to any
`selection' --- the numerator is supplied not by selection
but by Alice's existence.
To achieve the same result with SSA, Alice needs
to be coarse-grained, so the data is not `Alice was selected'
but `a Welsh person
was selected'. Such coarse-graining seems 
arbitrary and is in fact unnecessary under FIC --- the
same result can be obtained from the joint proposition
(Alice Exists, Alice Selected).

We have included being Welsh in the very definition
of Alice when writing down the $P({\rm AE})$ term.
We will continue to use this convention throughout all examples.
We
define Alice in the strictest possible terms, rather
than holding out particular properties (e.g. being Welsh)
for a Bayesian update. This procedure makes more sense
in many cases. Alice may have known she was Welsh
for as long as she can remember,
and we do not have to pretend that she forgot it and
then suddenly learned or remembered it as an adult in order to force
a Bayesian updating interpretation based solely on that
fact.

We note that the FIC likelihood based on (AE, AS)
is equivalent to the following procedure. Suppose you
are running an elaborate n-body simulation in which
beings (or some other objects of interest) may or may
not arise. If no such beings arise within the program,
the resulting output tells you that there were no beings.
If beings do arise in the program, one of them is selected
at random (i.e., with uniform probabilities)
and its details are returned to you. Those details
are the data from which you are to make inferences.

The likelihood formula (Equation~\ref{eqn:wal_likelihood1})
appears to involve a slightly complicated dependence on the two
unknowns $N_{\rm Wales}$ and $N_{\rm USA}$. However, it can be clarified
significantly by a coordinate transformation from
$(N_{\rm Wales}, N_{\rm USA})$ to $(N, f_{\rm Wales})$,
where
\begin{align}
N &= N_{\rm Wales} + N_{\rm USA}
\end{align}
is the total population of the two countries, and
\begin{align}
f_{\rm Wales} &= \frac{N_{\rm Wales}}{N}
\end{align}
is the proportion that are Welsh. Thus, the likelihood is simply
\begin{align}
P({\rm AE, AS}) &= f_{\rm Wales}.
\end{align}
The evidence of having been born in Wales increases
the plausibility of high values of the proportion $f_{\rm Wales}$, while (apparently) containing
no information about the total
population.

We stress that the data provides no evidence about the total population
only in the sense that the likelihood function does not
involve $N$. If the prior $p(f_{\rm Wales}, N)$ includes any dependence between $f_{\rm Wales}$ and $N$
then the data could still modify beliefs about $N$.
In applications of FIC (such as in Section~\ref{sec:m_stars}),
the data often provides evidence
about certain quantities despite them not appearing in the
likelihood expression at all, due to prior dependence
between the parameters. Prior dependence in FIC applications
is the norm, not the exception.

The likelihood expression, $f_{\rm Wales}$, does not imply that Alice now
believes Wales must definitely be larger than the USA.
In the scenario where Alice was raised isolated in a jail cell,
if she started with a Uniform$(0, 1)$
prior for $f_{\rm Wales}$ she would have a prior probability of $1/2$
for Wales being larger than the USA. After the update, the
posterior probability becomes $3/4$.
The generalisation to multiple countries is a Dirichlet$(1, 1, ..., 1)$
prior which gets updated to Dirichlet$(2, 1, ..., 1)$ (with Wales in the first slot without loss of generality) by Alice's
Welsh birth data. The posterior probability for Wales being larger
than the USA would still be $3/4$.

\begin{figure}
\begin{center}
\includegraphics[width=0.48\textwidth]{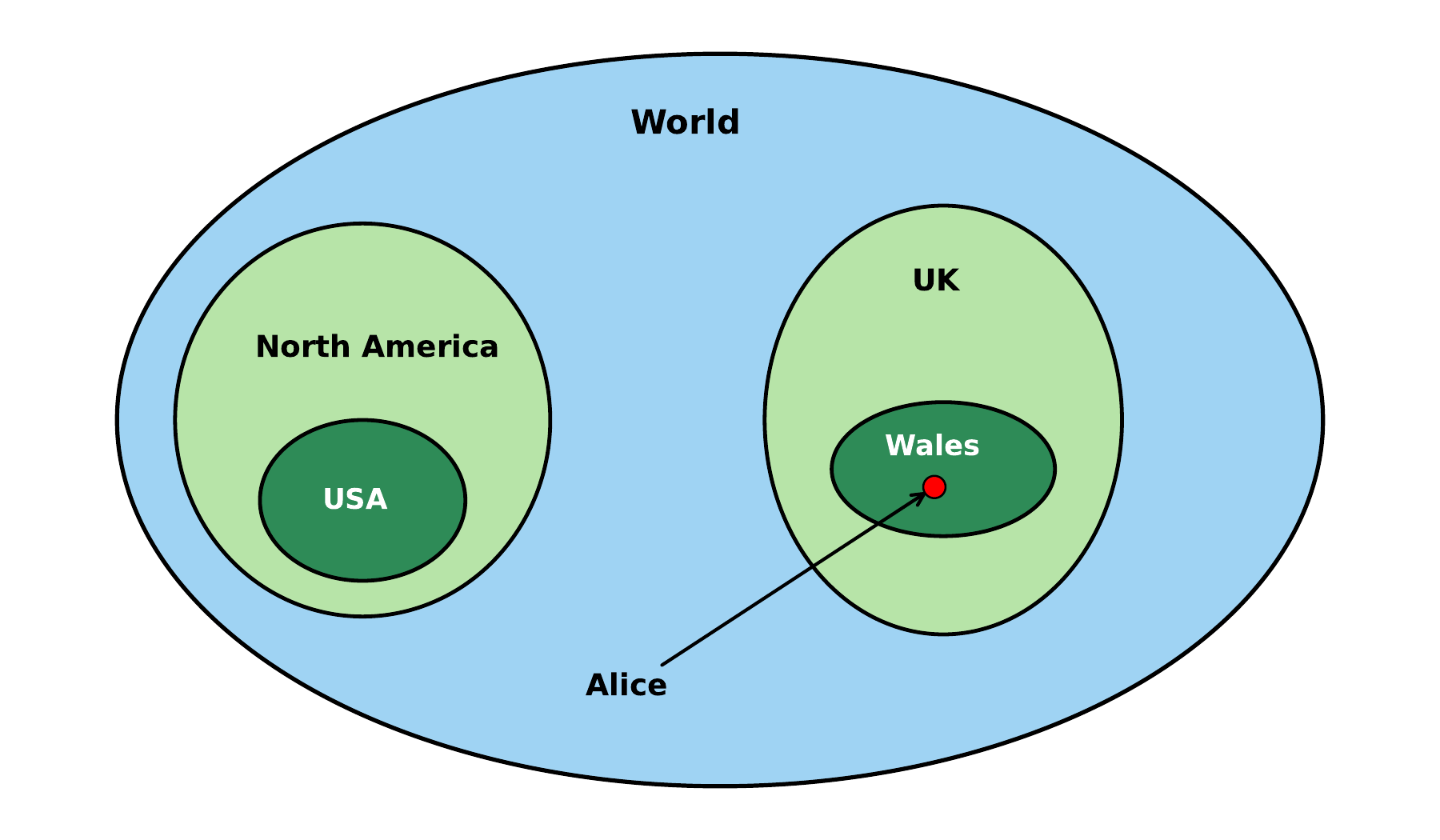}
\end{center}
\caption{A Venn diagram showing three sets that
Alice is included within, and some sets she is not.
Her evidence is relevant to the inference of the
relative sizes of nested sets she is in.
\label{fig:venn}
}
\end{figure}

In the analysis presented so far, the reference class is the union
of Wales and the USA.
Now, consider a modified version of the same problem, introducing a superset
of Wales called the United Kingdom, and a superset of USA called North America (shortened to NA in the equations).
A Venn diagram of the situation is shown in Figure~\ref{fig:venn}.
We can write down the FIC likelihood by using the fact that
each Welsh person has a probability of $\epsilon$ of being
Alice, but now considering Alice as having been selected
not from the union of Wales and the USA, but from a larger
reference class made up of the UK and North America:
\begin{align}
P({\rm AE, AS})
    &= P({\rm AE})P({\rm AS} \given {\rm AE}) \\
    &= \left(N_{\rm Wales}\epsilon\right)\left(\frac{1}{N_{\rm UK} + N_{\rm NA}}\right)\\
    &\propto \frac{N_{\rm Wales}}{N_{\rm UK} + N_{\rm NA}}.\label{eqn:wal_likelihood2}
\end{align}
This is similar to the original likelihood (Equation~\ref{eqn:wal_likelihood1})
except the denominator is different. Because the denominator is different,
it appears we will have
different conclusions --- a problem which has plagued SSA and similar ideas
such as FIC.

However, upon closer inspection,
the conclusions do not actually change. The new likelihood
(Equation~\ref{eqn:wal_likelihood2})
refers to different quantities from the original
one (Equation~\ref{eqn:wal_likelihood1}) and is therefore defined on a different
parameter space.
If we change coordinates on the respective parameter
spaces from population sizes to 
to proportions or
fractions, the apparent dependence on the reference class vanishes, and the
form of the FIC likelihood becomes clearer.

By the chain-product rule of probabilities, we can decompose the proportion
of the total population that is Welsh into a product of two terms;
the proportion of the total that is in the UK, multiplied by the proportion
of the UK that is Welsh:
\begin{align}
f({\rm Wales} \given {\rm Total}) &= f({\rm UK} \given {\rm Total})
                                 f({\rm Wales} \given {\rm UK}).
\end{align}
This is the same term that appears in the FIC likelihood with the
enlarged reference class (Equation~\ref{eqn:wal_likelihood2}),
but it has now been decomposed into a product of two other terms,
which may be unknown. Thus, the likelihood is simultaneously
proportional to all three terms, and inference for any of the
factors will involve a likelihood proportional to that factor.
See Appendix~\ref{sec:product_likelihoods} for a description
of how inference for only one factor works when treating the
others as nuisance parameters to be integrated out.

In general, FIC likelihoods will have the same structure
for any unknown proportion of the form
\begin{align}
f\left(\textnormal{some set Alice is in} \given
\textnormal{some larger set Alice is in}\right).
\end{align}
In the country population example, this includes
\begin{align}
f\left({\rm Wales} \given {\rm Wales}~\cup~{\rm USA}\right)
\end{align}
which is the result of the calculation from
the original reference class (Equation~\ref{eqn:wal_likelihood1}).
We conclude that the reference class problem is only an apparent
problem caused by using population sizes in the coordinate system.
A coordinate change to define unknowns as proportions,
to which the chain-product rule applies, shows that
conclusions do not actually depend on arbitrary reference classes.
In the population example, we could consider all of the
following as unknown parameters:
$f({\rm NA} \given {\rm World})$,
$f({\rm USA} \given {\rm NA})$,
$f({\rm UK} \given {\rm World})$,
$f({\rm Wales} \given {\rm UK})$,
with the constraint that all of these are between 0 and 1.
The likelihood is proportional to
\begin{align}
f({\rm Wales} \given {\rm UK})
f({\rm UK} \given {\rm World})
\end{align}
and also
\begin{align}
f({\rm Wales} \given {\rm Wales} \cup {\rm USA})
f({\rm Wales} \cup {\rm USA} \given {\rm World}),
\end{align}
depending on the choice of factorisation, which
depends on the scientific question.
Note that the priors for 
$f({\rm Wales} \given {\rm UK})$ and
$f({\rm UK} \given {\rm World})$ might reasonably be
independent, but for quantities such as
$f({\rm Wales} \given {\rm UK})$
and $f({\rm Wales} \given {\rm UK} \cup {\rm USA})$ the priors
should probably be dependent.

Ultimately, we could have the total population of the world
and the fraction in each country as free parameters, and the
likelihood would simply be $f_{\rm Wales}$,
with the data also being informative about other quantities
such as $f({\rm UK} \given {\rm World})$
through the mechanism of dependence in the priors.
This strongly resembles Bayesian density
estimation \citep{escobar1995bayesian}
from a single data point, along with the recognition
that the probability theory formulas such as the chain-product
rule are applicable to the density function $f(x)$
being estimated. If the scientific question involves definite
integrals of the density function over certain subsets of the
space, the analysis can either be done in terms of $f(x)$
with the integrals being treated as derived parameters,
or alternatively, the likelihood may be written down directly
in terms of the integrals of interest as we initially
did at the beginning of this Section.

In some cases (e.g., Sections~\ref{sec:abundance} and~\ref{sec:abiogenesis}),
the parameters being inferred are related not just to the specific
data but also to the fact that
the sample size is one, when it could have been zero.
When this is done, the FIC likelihood
usually involves a product of an existence term and proportion
terms.

It is worth noting that there are actually two reference classes
in the calculations we have just performed. The larger (the one traditionally
called a reference class)
is the set from which
Alice has been selected. The smaller is the finest-grained
proportion that is being inferred --- in this case, the proportion
of Welsh people. All other Welsh people would reach the same conclusions as Alice. In standard statistical terms, being
Welsh is a sufficient statistic for $f_{\rm Wales}$.
This implies that existing published calculations
that condition not on a detailed memory state but a coarse-grained one
(e.g., `I am a human observer') are acceptable, but are limiting
their conclusions to that resolution and above.
Conditioning on Alice's exact memories and using $\epsilon$
to do so is not strictly necessary to get the correct result for
coarser questions,
but is correct in principle and clarifies the underlying
reasoning process.

If we restrict the reference class from which Alice is selected
to be a smaller set than
the question of interest, FIC produces incorrect results.
For example, if Alice is interested in the proportion of
people in the UK that are Welsh, the reference class needs
to be at least as big as the UK.
If she uses a smaller reference class, for instance
only includes herself,
the denominator will disappear from the likelihood.
Without the denominator, FIC becomes FNC,
and can yield `presumptuous philosopher' conclusions analogous
to SIA.

FIC may also be extended to analyse the combined information
of multiple reasoners; see Appendix~\ref{sec:sharing} for
an illustration of this.
In the next section, we show that the Doomsday argument's
likelihood function is also given by a proportion
but with an extra factor due to learning of a rank within
the population.

\subsection{The Doomsday Argument}\label{sec:doomsday}
The Doomsday argument attempts to infer $N$, the number of humans
who will ever live, from $k$, one's observed birth rank
\citep{gott1993implications, bostrom2001doomsday}.
\citet{gott1993implications}'s analysis uses a classical confidence
interval style of reasoning, which effectively gives the same
conclusions as most Bayesian analyses --- that $N$ is probably
not greater than $k$ by a huge factor.
For people
alive today, $k \approx 5\times10^{10}$,
but the exact number can vary depending
on the inclusiveness of the definition of `human' in prehistory
as well as changes in paleontological evidence.
Initially, we will assume the total population $N$ and the birth rank $k$
refers to \sapiens.
As in the country example, we will consider the definition
of `Alice exists' to include all of her details, in this case including
her knowledge of her birth
rank $k$, which is not treated as a separate piece of information.
This is to emphasise that the FIC likelihood is, in principle,
only $P({\rm AE}, {\rm AS})$, using the most precise possible
definition of what it means for Alice to exist.
The FIC likelihood is given by
\begin{align}
P({\rm AE}, {\rm AS})
    &= P({\rm AE})P({\rm AS} \given {\rm AE}) \\
    &=     \left(\epsilon\mathds{1}(N \geq k)\right) \times \left(\frac{1}{N}\right),
\end{align}
where $\mathds{1}()$ is an indicator function that is 1
if the argument is true and 0 if it is false.
The probability for Alice to exist is based on the
person at rank $k$ having Alice's other properties (with probability
$\epsilon$), with the truncation arising from the fact this
is only possible at all if $N \geq k$, i.e., that a person
with birth rank
$k$ even exists.
The overall result here agrees with the traditional
SSA conclusion and mildly
downweights very high possibilities for $N$.
In the previous section, we observed that the
FIC likelihoods were unknown fractions.
$1/N$ is in fact a fraction --- the fraction of
\sapiens~that is Alice.

Importantly, there is no SIA-style term proportional to $N$.
To get such a term to appear, the definition of Alice would need
to exclude her birth rank, so that there are $N$ slots where
she might appear. However, this extra $N$ term, if introduced,
would immediately be
cancelled by a {\em second} $(1/N)$ factor from the datum of
her birth rank $k$.

To clarify the nature of this inference and the reference class
problem, we can divide the timeline into various sets and apply the
same logic as in the country population example. One simple difference here
is that the sets are defined temporally instead of spatially.
Another difference is that Alice's birth rank $k$ locates her
at a specific point within \sapiens, not just somewhere within that
subset. In the country problem, Welsh people were not arranged in some order
with Alice placed at a particular known location within that ordering.
This extra ordering-related data explains why the result here
gives some leverage about the total population size, whereas
for the country problem the likelihood did not involve $N$.
Intuitively, in the Doomsday
problem, $N$ and $k$ share a common scale, so learning $k$
is slightly informative about $N$.

We now consider a more complex model involving the genus \homo~and the
species \sapiens, which is a subset of \homo.
Within the \sapiens~subset sits
Alice, with her birth rank of $k$ (defined with respect to \sapiens).
We will now attach a subscript $S$ to the total population of
\sapiens,
which is now $N_S$. $N_H$ will represent the total membership of the genus
\homo, where $N_H > N_S$.

We can also define the fraction of \homo~that is \sapiens,
as
\begin{align}
f &= f({\rm Sapiens} \given {\rm Homo}) = \frac{N_S}{N_H}.
\end{align}
By the chain-product rule we have
\begin{align}
f({\rm Alice} \given {\rm Homo})
    &= f({\rm Sapiens} \given {\rm Homo})f({\rm Alice} \given {\rm Sapiens}) \\
    &= f({\rm Sapiens} \given {\rm Homo}) \times \frac{1}{N_S} \\
    &= \frac{1}{N_H}.
\end{align}
which is the same as the original `Sapiens-only'
FIC likelihood but multiplied by another unknown proportion.
This shows that Alice's evidence also tells her not just about the size
of \sapiens~($N_S$),
but also about the fraction of \homo~that is \sapiens,
and also the size of \homo~in total ($N_H$).

Thus, Alice can simultaneously perform two overlapping Doomsday analyses,
one for \sapiens~and another for \homo. Both involve a likelihood proportional
to $1/N$ for their respective total populations. Note that her birth rank
$k$, which only provides a truncation (if it is known),
would be larger in the broader \homo~analysis.
We now perform the analysis explicitly in the case where $k$ is
only known with respect to \sapiens,
and Alice is `selected' only from \sapiens:
\begin{align}
P({\rm AE}, {\rm AS})
    &=     \left(\epsilon\mathds{1}(N_S \geq k)\right) \times \left(\frac{1}{N_S}\right) \\
    &= \left(\epsilon\mathds{1}(fN_H \geq k)\right) \times \left(\frac{1}{fN_H}\right) \\
    &= \left\{
        \begin{array}{lr}
        \frac{1}{fN_H}, & fN_H \geq k \\
        0, & \textnormal{otherwise}.
        \end{array}
        \right.
\end{align}
If Alice is selected from \homo~instead of \sapiens, we get
\begin{align}
P({\rm AE}, {\rm AS})
    &=     \left(\epsilon\mathds{1}(N_S \geq k)\right) \times \left(\frac{1}{N_H}\right) \\
    &= \left(\epsilon\mathds{1}(fN_H \geq k)\right) \times \left(\frac{1}{N_H}\right) \\
    &= \left\{
        \begin{array}{lr}
        \frac{1}{N_H}, & fN_H \geq k \\
        0, & \textnormal{otherwise}.
        \end{array}
        \right.
\end{align}
The difference between the two expressions is a constant factor
$(1/f)$.
For inference about $N_H$, this
irrelevant proportionality constant. However, if $N_S$
and/or $f$ is of interest, the factor $f$ is relevant
and can be retained. Note that in principle, nuisance
parameters such as $f$ should be assigned a prior and
marginalised over as discussed in Section~\ref{sec:product_likelihoods}.

The analysis could be broadened further to include all mammals, all animals,
etc, a set of nested reference classes, or to include known
birth ranks with respect to each class. Each conclusion would involve a
$1/N$ factor in the likelihood, but a different truncation position due to the
differing birth ranks (if they are known).
Broader reference classes will lead to larger birth ranks
and hence more `optimistic' future predictions
for that reference class.

\subsection{Edge Cases}\label{sec:edge_cases}
Authors such as \citet{neal2006puzzles} have argued forcefully against
the $1/N$ factor in the Doomsday likelihood,
preferring a likelihood function that merely truncates
the possibilities to $N \geq k$.
Neal uses FNC,
which has the existence term but not the sampling term, so
the derivation is equivalent to ours but with one fewer
$1/N$ factor.
One of the objections to the $1/N$ dependence is that
some people within the \sapiens~timeline would make
peculiar inferences. 

For example, if a worldwide calamity occurs
and Alice discovers she is the last human alive,
she can perform a time-reversed Doomsday-style
inference based on her `death rank' $k' = 1$.
The $1/N$ likelihood penalises hypotheses where the total human population
in the past was large. According to FIC, this is a correct
conclusion. However, in a more realistic version of this
scenario, it is unlikely Alice could ever actually learn
that $k' = 1$ as it would require global knowledge of the whole
Earth to rule out any other survivors.
More realistically, Alice's survival (which is trivially
knowable in practice)
would actually
provide some
evidence that the proportion of humanity that survived
is large. Moreover, in the calamity situation, Alice would
have much more information about the past population
than current humans have about the future population.
Stronger prior distributions are less prone to being
pushed around by a relatively modest $1/N$ likelihood factor.

Similar issues occur in the country population example.
For example, people born in Monaco would
have a likelihood factor proportional
to $f_{\rm Monaco}$ for the proportion of the world's population
born in Monaco (which is around $4 \times 10^{-6}$ in reality).
In the absence of strong
corrective evidence, this could lead them to assign somewhat
high probability
to incorrect statements about the proportion of people in
various countries, e.g., with a flat Dirichlet prior,
$P(f_{\rm Monaco} > f_{\rm USA} \given {\rm Monaco~birth}) = 3/4$.

There is nothing mysterious about these results in terms
of statistical inference.
All ordinary, non-anthropic Bayesian models include the possibility
of observing tail datasets that give a misleading posterior
distribution.
For example, suppose some data $\{x_i\}_{i=1}^{100}$
has a normal sampling distribution with mean $\mu$ and
standard deviation $\sigma$,
and we wish to infer $\mu$ and $\sigma$ from the data.
If the true values are $\mu=0$ and $\sigma=1$,
it is entirely possible to obtain data whose arithmetic mean
is $x=10$ and whose
standard deviation is $5$. The joint posterior distribution
for $(\mu, \sigma)$ would be right in the sense that it
was calculated correctly, but wrong in the sense
that the true values of $\mu$ and $\sigma$
would be far in the tails of the
posterior distribution.
There has never been a guarantee in Bayesian inference
that all possible datasets will give `good'
posterior distributions in the sense of assigning
a high probability to the true values of the unknown parameters.
The primary reassurance comes from
the fact that these
misleading datasets have a low prior probability according to the prior predictive distribution
$p(x_1, ..., x_N) = \iint p(\mu, \sigma)p(x_1, ..., x_N \given \mu, \sigma) \, d\mu \, d\sigma$.
Identifying these misleading datasets
does not refute the model. However,
they may provide a check on the model assumptions.
If you think the model has assigned these datasets too little
prior probability, you should change something about the model.
A harder guarantee from Bayesian updating,
that is true for all possible datasets,
is discussed in Appendix~\ref{sec:guarantee}
for completeness.

While ideas about inferential procedures producing good
results for a majority of possible datasets is traditionally
associated with frequentist inference, these ideas are present in
Bayesian inference also. Indeed, any Bayesian model with
parameters and data can be factorised in the traditional
way ($p(\theta)p(D \given \theta)$) or in reverse
($p(D)p(\theta \given D)$). The reverse way describes
the model in terms of a probability distribution over
datasets and a mapping from data to posterior distributions.
On the other hand, frequentist concepts
regarding reliability of inferences for most datasets
only involve
the sampling distribution $p(D \given \theta)$, strictly
interpreted as a model of relative frequencies.
We note that one distinction between ordinary Bayesian
inference and anthropic problems is that, in the former,
alternative datasets are hypothetical and are only considered
in the prior state of knowledge. In anthropic problems,
the alternative possible datasets are actually realised ---
for example, in the memories of people born in other countries
or with different birth ranks.

\section{Astrobiology Applications}\label{sec:astrobiology}

\subsection{Application I: The Abundance of Life}\label{sec:abundance}
Consider the problem of estimating the abundance of life
and conscious observers in the universe from your own existence.
It has been argued \citep[e.g.][]{carter1983anthropic}
that our existence hardly provides any evidence about the
abundance of life in the universe.
More evidence than just our own existence can be considered
\citep{frank2016new, sandberg2018dissolving}, but we do
not consider it here.
This problem can be treated with FIC and confirms this
standard argument about the uninformativeness of our
existence, with the additional conclusion
that the universe needs to be such that the probability of at
least one observer is moderate.
Application of FIC in this problem is complicated by the
fact that the number of observers at each suitable `site'
(e.g. habitable planet) in the universe needs to
be modelled explicitly and then marginalised over
in order to compute the $P({\rm AS} \given {\rm AE})$
term. It is possible in principle to write this term down
directly, but it is harder to get the resulting expression right.

Suppose the universe contains $N$ suitable sites where life
could arise. At each site, the probability of life arising
is $\theta$, and the conditional probability of conscious observers
arising given life has arisen is $\phi$.
At any given site, the probability that Alice exists at that
site is then $\theta\phi\epsilon$, where $\epsilon$
is a very small probability encapsulating all of Alice's specific
properties. Therefore, the probability that Alice exists at
any site is
\begin{align}
P({\rm AE}) &= 1 - \left(1 - \theta\phi\epsilon\right)^N.
\end{align}
Since $\epsilon$ is very small this is approximately (by Taylor expansion
around $\epsilon=0$)
\begin{align}
P({\rm AE}) &\approx N \theta \phi \epsilon.
\end{align}
On its own, this would be the likelihood under FNC.
However, it yields very strong conclusions in favour of
abundant life.
To see this,
suppose $\theta$ and $\phi$
are initially assigned log-uniform priors independently, so
$p(\theta, \phi) \propto 1/(\theta\phi)$ between some small lower limit
and 1. The FNC likelihood
cancels this out and we are left with uniform posteriors
independently for $\theta$ and $\phi$, between a small lower
limit and 1.
Thus, under FNC, merely from her own existence (even isolated in a jail cell as in the country population example), Alice concludes
that with a probability of 99.9\% that $\theta$ is greater than 0.001,
and therefore there are almost certainly a large number of life-hosting
planets in the Milky Way.
This is the `Presumptuous Philosopher' problem, here resulting from FNC
rather than SIA.
Once we multiply the likelihood by $P({\rm AS} \given {\rm AE})$
to implement FIC, this conclusion will vanish.

\begin{figure}
\begin{center}
\includegraphics[width=0.48\textwidth]{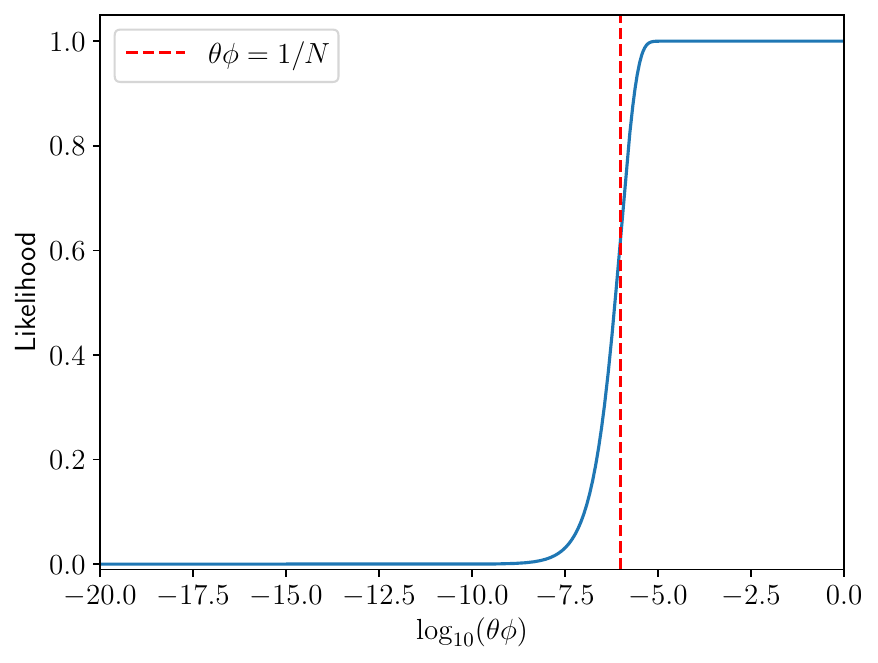}
\end{center}
\caption{The likelihood function for the probability of
observers arising at a given site $(\theta\phi)$ from
Alice's data. The data suggests that $\theta\phi$ is
greater than about $1/N$. Here, the number of sites
is $N=10^6$ for illustration.
\label{fig:abundance}
}
\end{figure}

To compute this extra factor and calculate the FIC likelihood,
we will consider the reference class to be conscious observers.
Our definition is intentionally vague. However, as discussed
in Section~\ref{sec:countries}, changes in the reference class
are unlikely to modify the result.
Denote the number of conscious observers at each site by
$\{m_1, ..., m_N\}$ and let the total be $M = \sum_i m_i$.
If we knew $M$, the FIC likelihood would be
\begin{align}
P({\rm AE}, {\rm AS}  \given M)
    &= P({\rm AE} \given M)P({\rm AS}  \given {\rm AE}, M) \\
    &= \left\{
        \begin{array}{lr}
        (\epsilon M)(1/M), & M \geq 1 \\
        0,         & \textnormal{otherwise.}
        \end{array}
        \right. \\
    &= \left\{
        \begin{array}{lr}
        \epsilon, & M \geq 1 \\
        0,         & \textnormal{otherwise.}
        \end{array}
        \right.
\end{align}
However, we need to eliminate the dependence on $M$ by
marginalising over it:
\begin{align}
P({\rm AE}, {\rm AS})
    &= \sum_{X=0}^\infty P(M)P({\rm AE}, {\rm AS}  \given M) \\
    &=  \sum_{X=0}^\infty P(M) \epsilon \mathds{1}(M \geq 1) \\
    &= \epsilon P(M \geq 1).
\end{align}
This is equal to the probability that at least one observer
arises somewhere, multiplied by the very small probability
for any observer to be Alice in detail, i.e.,
\begin{align}
P({\rm AE}, {\rm AS})   &= \epsilon\left( 1 - (1 - \theta\phi)^N\right).
\end{align}
Therefore, merely from her own existence, Alice does not learn much about $\theta$
and $\phi$, but her existence supports values where $\theta\phi \gtrsim (1/N)$,
as shown in Figure~\ref{fig:abundance}.
Alice's data only rules out parameter values where {\em any}
observer would be unlikely to exist.
For illustrative purposes we take $N=10^6$, even though the true
number of habitable planets is potentially much larger
\citep{petigura2013prevalence, dressing2013occurrence}.

Unlike in the previous examples, this likelihood is not a proportion,
but is a modest form describing the evidence provided by the fact
that at least one observer exists. The scientific question did not
involve any of Alice's details and the only information
comes from the fact that the sample size is not zero.

\subsection{Application II: The Early Abiogenesis Time}\label{sec:abiogenesis}
The fact that life apparently arose on the Earth soon after its
formation may be taken as evidence that
abiogenesis is not difficult\footnote{While panspermia might be a possibility, we
will not consider it in this section.}.
Exactly how strong the evidence is has been a matter of debate.
Several authors have studied the evidence with formal probabilistic
models, beginning with \citet{lineweaver2002does}.
The parameter of interest can be considered to be the intensity
$\lambda$ of a Poisson process, with abiogenesis being the
occurrence of the first event in the Poisson process.
Presumably $\lambda$ is a function
of the physical and chemical conditions on a planet, but for simplicity
we will not be considering this dependence
but will treat $\lambda$ as simply an unknown constant, measured
in units of (Gyr)$^{-1}$, that applies on each of $N$ planets
deemed habitable.
\citet{lineweaver2002does} found strong evidence for a large
value of $\lambda$ (although they did not parameterise it in these
terms). However, \citet{brewer2008implications} found that this
conclusion was largely driven by an unintentional
informative prior. Switching to a log-uniform prior
for $\lambda$ over many orders of magnitude, \citet{brewer2008implications}
concluded that while there is weak evidence in favour of higher
values of $\lambda$, the likelihood function is close to a nonzero constant at moderately
low values, which remain very much in play.
This type of argument was further refined, taking into account more physical
considerations, by \citet{spiegel2012bayesian} (who noted the same problem
with Lineweaver and Davis's informative prior)
and \citet{kipping2020objective}.

Moreover, different authors have written down different forms
for the likelihood function for updating knowledge about $\lambda$
from observation of $t_a$, the abiogenesis time.
Following \citet{lineweaver2002does} we take $t_a$ to be 0.5 Gyr.
We also assume abiogenesis could have occurred at any time
from $t=0$ onwards, despite the fact that physical processes
such as Late Heavy Bombardment \citep{gomes2005origin}
could have prevented this.
\citet{brewer2008implications}'s likelihood is of the form
\begin{align}
L_1(\lambda) &= \frac{\lambda\exp(-\lambda t_a)}{1 - \exp(-\lambda t_{\rm now})},\label{eqn:brewer_likelihood}
\end{align}
where $t_{\rm now}$ is the current time with respect to the
Earth's origin (about 4.5 Gyr).
The argument is that abiogenesis has to be observed at an
earlier time than the observer is present, which truncates
the exponential distribution from the Poisson process
to $[0, t_{\rm now}]$ (the denominator
in the expression comes from normalising the truncated exponential
distribution).
However, \citet{kipping2020objective} wrote down a different likelihood
function. His model is more complex, so we present a simplified
version of it here for comparison with \citet{brewer2008implications}.
Instead of an exponential distribution normalised over
the interval $[0, t_{\rm now}]$, \citet{kipping2020objective}'s
likelihood function involves a window during which abiogenesis
could happen, which may be a different interval of time.
For simplicity we use the 
the interval $[0, T]$:
\begin{align}
L_2(\lambda) &= \frac{\lambda\exp(-\lambda t_a)}{1 - \exp(-\lambda T)}.\label{eqn:kipping_likelihood}
\end{align}
A simple understanding would set $T$ to the main sequence
lifetime of the Sun
(about 10 Gyr), but it could also be significantly shorter based on
terrestrial habitability
considerations \citep[e.g.][]{falkowski2005rise}.

We derive the FIC result for this simple model of the early
abiogenesis situation in Appendix~\ref{sec:abiogenesis_derivation}.
From the result, we will see that
Equations~\ref{eqn:brewer_likelihood} and~\ref{eqn:kipping_likelihood}
are actually completely compatible, differing only in which
parts of the total information are considered prior information and
which parts are considered data in the sense of Bayesian updating.
Moreover, our result will include an additional factor
analogous to Section~\ref{sec:abundance}, regarding the evidence
about $\lambda$ that is provided by the fact that
anybody exists at all.

\begin{figure}
\begin{center}
\includegraphics[width=0.48\textwidth]{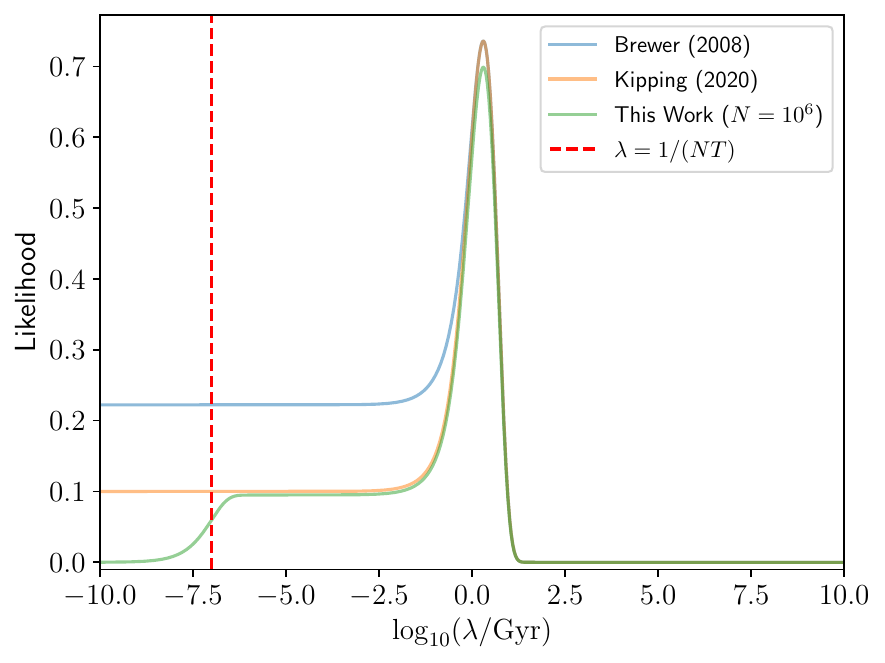}
\end{center}
\caption{The three likelihood functions for the abiogenesis
rate parameter $\lambda$.
None of the likelihoods are incorrect, but they refer to different
subsets of the available information. The likelihood in this work
is for the data of existence, the current time, and the abiogenesis
time. The \citet{brewer2008implications} likelihood is for just the abiogenesis time,
with existence and the current time factored into the prior.
The \citet{kipping2020objective}-style likelihood
(with the denominator involving $T$ rather than $t_{\rm now}$)
is for the abiogenesis time and the current
time, with existence factored into the prior.
For illustration purposes $N$ is set to $10^{6}$, even though
the actual number of habitable planets in the universe is plausibly
larger.
\label{fig:abiogenesis}
}
\end{figure}

The result of the FIC derivation is
\begin{align}
P({\rm AE}, {\rm AS}) &\propto \left(1 - \exp(-N\lambda T)\right)
\left(\frac{\lambda\exp(-\lambda t_a)}{1 - \exp(-\lambda T)}\right).
\end{align}
The second factor can also be expressed as the product of two distinct
factors. The three factors and their interpretations are given below:
\begin{align}
P({\rm AE}, {\rm AS}) &\propto
\overbrace{\left(1 - \exp(-N\lambda T)\right)}^{\textnormal{Existence of anybody}} \times
\overbrace{\left(\frac{1 - \exp(-\lambda t_{\rm now})}{1 - \exp(-\lambda T)}\right)}^{\textnormal{Alice at }t_{\rm now}} \nonumber\\
& \quad\quad\quad\quad\quad \times
\overbrace{\left(\frac{\lambda\exp(-\lambda t_a)}{1 - \exp(-\lambda t_{\rm now})}\right)}^{\textnormal{Abiogenesis at }t_a}.
\end{align}
The two different likelihoods from the literature are both contained within
this expression. \citet{brewer2008implications} presented just the final
factor, which is relevant for updating on $t_a$ alone.
The simplified version of
\citet{kipping2020objective}'s likelihood includes the final two
factors, and is appropriate for updating when given both
$t_a$ and $t_{\rm now}$.
This work adds the additional, weak factor for the existence of
any observers at all, which is analogous to the likelihood
derived in Section~\ref{sec:abundance} but written in terms
of $\lambda$.
The other terms can be interpreted as proportions, as in the population
and Doomsday examples. Specifically, they represent the proportion
of all observers with whom Alice shares her values of $t_a$
and $t_{\rm now}$.
The three likelihood functions are displayed in Figure~\ref{fig:abiogenesis} in terms of $\log_{10}(\lambda)$.

Note that, in this section, observers and life are not treated
distinctly as in Section~\ref{sec:abundance}, and we are vague
about their definitions.
More complex models
are possible; indeed, \citet{kipping2020objective} includes a
second Poisson process with an unknown rate parameter for
the possible development of life into intelligence.
The fact that we appear close to the end of a terrestrial
window $[0, T]$ provides weak evidence that this second
Poisson process has a low value for its rate parameter,
an argument also advanced by \citet{snyder2021timing}.
The ideas presented in this paper suggest that such conclusions
may not be sensitive to any particular definition of intelligence
(or any other category that we might place ourselves in), but
likely apply regardless of whether the adopted criteria
are narrow or broad.

\subsection{Application III: The Red Sky Paradox and the Cosmic Time}\label{sec:m_stars}
The fact that we find ourselves around a G type star,
even though M dwarf stars are more common (and will be more so
in the future of the universe), can be taken as suggesting that
G star life might be
a moderate to large proportion of all life, and therefore the
per-star probability of developing life around an M dwarf is low.
\citet{kipping2021formulation, kipping2026solar} developed this argument formally
using Bayesian inference and a realistic treatment of the
astronomical factors, particularly considering
the expected much greater number of M dwarf-years in the future
of the universe.

Here, we discuss a simple formulation of this
problem to discuss the inference
in terms of FIC to clarify some questions surrounding it.
In a simple model, the likelihood function for the number of
occurrences of G star and M star
life, from the fact we are around a G star, is given by:
\begin{align}
L(N_G, N_M) &\propto \frac{N_G}{N_M + N_G},\label{eqn:m_star_likelihood}
\end{align}
which is equivalent to
the country population example (Section~\ref{sec:countries}).
For simplicity, our analysis considers the ratio of the total
populations,
rather than per-star probabilities, with consideration of the greater
absolute number of M stars applied informally outside of the modelling.

This formula is traditionally obtained through SSA-style arguments,
although, as in Section~\ref{sec:countries},
under FIC only the denominator results from sampling,
with the numerator relating to existence.
However, it is unclear whether $N_G$ and $N_M$
are the numbers of individual observers (reference class members)
around each star type,
the number of life-bearing planets around each star type,
or the number of civilisations, however defined,
around each star type. We argue that it is the former ---
the number of individuals around each star type.
The reason is due to the mechanism by which the universe
appears to produce individuals who are aware of their own
memories and circumstances. Processes occur in the universe
which lead to individuals and each individual possesses their
own personal collection of information.

Let $\{m_i\}$ be the population of reference
class members around each G star and
$\{n_i\}$ be the populations around each M
star. The probability for Alice to exist and be selected is
\begin{align}
P({\rm AE}, {\rm AS})
    &= P({\rm AE})P({\rm AS} \given {\rm AE})
    &= \left(\epsilon \sum_i m_i\right)\left(\frac{1}{\sum_i m_i + \sum_j n_j}\right).
\end{align}
This is equivalent to Equation~\ref{eqn:m_star_likelihood}
once we identify $N_G = \sum_i m_i$ and $N_M = \sum_j n_j$
as the population totals.
If we want to know about the proportion of civilisations
(however defined) around each star type, these appear in
the likelihood expression as the length of each of the sums.
However, the unknown populations would need to be marginalised
over. With an assumption of equal prior expectations of
population size for G and M star civilisations, the familiar
formula can be obtained. However, this might be unrealistic.
More generally, Alice's data has been `sampled'
from a population-weighted probability distribution over
civilisations --- the fact we exist around a G star might simply
indicate that G stars produce more individuals, rather than
more civilisations.

We can also combine our observation of being around a G star
with the related knowledge of the current time $t=13.8$ Gyr
with respect to the Big Bang, which can provide
modest evidence about the temporal distribution of life
in the universe.
Suppose G stars produce life a total of $N_G$
times, with temporal density proportional to
a lognormal distribution with median $\mu_G$
and standard deviation in log space of $\sigma_G$.
Similar quantities $N_M$, $\mu_M$, and $\sigma_M$
may be defined for M stars. The likelihood function
is straightforward:
\begin{align}
P({\rm AE}, {\rm AS})
    &= \left(\epsilon N_G f_G(13.8~{\textnormal{Gyr}})\right)\left(\frac{1}{N_G + N_M}\right),\label{eqn:G_and_t}
\end{align}
where $f_G(t)$ is the probability density function for a lognormal
distribution. With the explicit functional form of a lognormal
density, the likelihood becomes
\begin{align}
P({\rm AE}, {\rm AS})
    &\propto \frac{\frac{N_G}{\sigma_G}\exp\left(-\frac{1}{2\sigma_G^2}(\ln(13.8/\textnormal{Gyr}) - \ln(\mu_G))^2\right)}{N_G + N_M},
\end{align}
where the $1/t$ factor from the lognormal distribution does not
involve the parameters and is absorbed into the proportionality
sign.
Thus, Alice's data provides a little bit of information
about the ratio of G star observers to M star observers
(supporting the conclusion that life around M stars might be
less common),
and some information about the temporal density of G
star observers (i.e., the parameters $\mu_G$ and $\sigma_G$).
The likelihood function does not involve
the temporal
density of M star observers. We may be tempted to say that
the data provides no evidence about the temporal density of
M star observers, but such a conclusion would require an
additional assumption of prior independence between
$(\mu_G, \sigma_G)$ and $(\mu_M, \sigma_M)$.
If the priors are quite dependent, then Alice's data
does provide a little evidence that M star observers are concentrated earlier in the universe's history.
If the priors are independent,
then the temporal distribution of M star observers
could still be largely in the future \citep{loeb2016relative}.

Interestingly, the likelihood based on the joint data
of the G star observation and $t=13.8$~Gyr may be factorised
differently, with the temporal density of both G and M-star
observers written first, multiplied by the conditional probability
of the G star fact given the current time.
This would appear
to provide evidence (in the sense of it appearing in the
likelihood function) about the temporal density of all observers
via the first term, but this is immediately cancelled by the
denominator in the second term. 
The easiest way to see this
is to rearrange Equation~\ref{eqn:G_and_t} into a form where it
is the product of a $t=13.8$~Gyr term and a G-star term:
\begin{align}
P({\rm AE}, {\rm AS})
    &\propto\overbrace{\left(\frac{N_G f_G(t) + N_M f_M(t)}{N_G f_G(t) + N_M f_M(t)}\right) }^{{\rm One}} \overbrace{\left(\frac{N_G f_G(t)}{N_G + N_M}\right)}^{\textnormal{Eqn\mbox{~\ref{eqn:G_and_t}}}} \\
    &= \overbrace{\left(\frac{N_G f_G(t) + N_M f_M(t)}{N_G + N_M}\right)}^{\textnormal{Being at time}~t}
    \overbrace{\left(\frac{N_G f_G(t)}{N_G f_G(t) + N_M f_M(t)}\right)}^{\textnormal{Being around a G star given time}~t}.
\end{align}
While $f_M(t)$ appears in the first term, it is immediately
cancelled by the second term's denominator.

Overall, the observation that we exist around a G star
disfavours hypotheses where M stars host a large majority
of the universe's life. However, the total amount of life
around M stars could still be comparable to the total amount
around G stars. If there is M star life in the universe,
its temporal distribution is potentially weakly informed by the observation
that we are at $t=13.8$ Gyr, depending on whether
we consider G star data to be relevant to M star parameters
(i.e., whether the priors are dependent).
Dependent priors would not be unusual here if there are
shared mechanisms for the development of life that apply
across stellar types.

Note that our arguments here are purely based on our own local observations
and do not take into account physical considerations about
the potential habitability of planets orbiting M dwarfs
\citep{tarter2007reappraisal}.

\section{Conclusion}\label{sec:conclusion}
In this paper we demonstrated a formal approach
to anthropic reasoning in astrobiology called Full Indexical
Conditioning (FIC), inspired by \citet{neal2006puzzles}'s Full
Non-Indexical Conditioning (FNC).
FIC is used within ordinary Bayesian
inference and clarifies the origins of the likelihood functions
required for making inferences from the data of our particular
circumstances.
FIC is closely related to the self sampling assumption (SSA) which
is often used in these kinds of models, but has a clearer
philosophical foundation. The key proposition that should be
conditioned on under FIC is the existence of the particular
observer doing the reasoning, as well as their `selection'
from a pool of potential observers whose data could have been
analysed but was not. It is this selection process that
distinguishes FIC from FNC.

Through examples, we demonstrated that
the reference class problems commonly
associated with FIC and SSA are an artifact of
writing the unknown parameters in terms of population
sizes instead of proportions. When proportions are used,
the chain-product rule of probabilities shows that the evidence
is relevant to many different proportions simultaneously.
This can occur even when certain unknown quantities do
not explicitly appear in the likelihood function, through
the mechanism of dependent priors.

When applied to astrobiological problems such as the abundance
of life in the universe,
FIC reproduces the standard conclusion, that our existence
is not highly informative about life in the universe except
for ruling out hypotheses where {\em any} observers would be
unlikely to arise.
We also re-derived known results in the literature regarding
the early abiogenesis time from an FIC foundation. The results
show that FIC provides an additional weak factor that was not
present in previous analyses which were implicitly performed
using SSA. However, this derivation required an assumption
that population sizes on each habitable planet are independent
of the abiogenesis time, which may be unrealistic.

Finally, we applied FIC to the question of the abundance
of life around M stars as compared to G stars, obtaining
the standard conclusion in agreement with SSA, and showing
that our observation time of $t=13.8$ Gyr may or may not
be informative about the temporal distribution of M star life,
depending on the choice of prior distribution.

In conclusion, FIC has the potential to
provide a clear and unified framework
for performing inference in situations involving anthropic
selection effects.

\section*{Acknowledgements}
It is a pleasure to thank Radford Neal (Toronto) for useful feedback on an inquiry
about his thought provoking paper,
which was the direct inspiration for this work (even though
we disagreed with it in the end).
We also thank Thomas Lumley (Auckland) for a helpful conversation
about edge cases in Bayesian inference.

Google Gemini Pro was used as a discussion partner for the ideas presented in this
paper. It did not contribute to the text.

\bibliographystyle{mnras}
\bibliography{anthropic} 

\appendix

\section{Product Likelihoods}\label{sec:product_likelihoods}
This paper addresses many problems where there are multiple unknown
parameters $x$, $y$, $z$, and the likelihood function is proportional
to their product $xyz$ (here we present the case with three
parameters; the generalisation to any number is straightforward).
In this case, Bayes' theorem gives the
joint posterior
\begin{align}
p(x, y, z \given D) &\propto p(x, y, z) xyz
\end{align}
If the prior is independent and can be factorised as
\begin{align}
p(x, y, z) &= p(x)p(y)p(z),
\end{align}
then the inference for each unknown parameter can be
performed separately, giving
\begin{align}
p(z \given D) &\propto p(z) z,
\end{align}
and similarly for $x$ and $y$.

However, if the prior does not factorise,
we can still integrate out
$x$ and $y$ to get the marginal posterior for $z$,
getting a related but more general result:
\begin{align}
p(z \given D) &= \int p(x, y, z \given D)\, dx \, dy \\
    &\propto \int p(x, y, z) xyz\, dx \, dy \\
    &= \int p(z)p(x, y \given z) x y z \, dx \, dy \\
    &= p(z) z \int p(x, y \given z) x y \, dx \, dy \\
    &= p(z) z \mathds{E} \left(xy \given z\right).
\end{align}
Thus, if only $z$ is of interest, the likelihood remains
proportional to $z$ but also includes an extra
term to account for whether particular values of $z$
have any effect (in expectation) on the product $xy$.

\section{Information Sharing}\label{sec:sharing}
Suppose Alice meets Bob, and wishes to perform inferences by
combining their information together. In this situation,
it is (presumably) known that both Alice and Bob exist.
Moreover, an inferential procedure to combine information from
two reasoners should work for any pair of reasoners.
Thus, we propose that the appropriate FIC likelihood
is given by the probability that Alice and Bob exist
and are selected as a pair:
\begin{align}
P({\rm AE}, {\rm BE}, {\rm AS}, {\rm BS}).
\end{align}
For example, suppose Alice's hair is brown and
Bob's is black, and they wish to infer the
proportions $f_{\rm brown}$ and $f_{\rm black}$
in the population, whose total size is $N$.
The likelihood function would be
\begin{align}
P({\rm AE}, {\rm BE}, {\rm AS}, {\rm BS})
    &= P({\rm AE}, {\rm BE})P({\rm AS}, {\rm BS} \given {\rm AE}, {\rm BE}) \\
    &= \left(N_{\rm brown}\epsilon_{\rm Alice}N_{\rm black}\epsilon_{\rm Bob}\right)
        \frac{1}{N^2} \\
    &\propto f_{\rm brown}f_{\rm black}.
\end{align}
This is equivalent to having selected one brown ball and
one black ball from an urn in a traditional multinomial statistical
setup\footnote{We have assumed sampling with replacement
as the formulas are simpler --- if the total population is
large enough it makes no difference.}.
In practice, if the computational philosophy professor knows
that the probability of Alice and Bob being the two people to
use the inference program
is greater than $1/N^2$, the result should be modified.
This could apply, to some degree, to hair colours because they are correlated for biogeographic
reasons and thus a typical person is more likely to
encounter people with the same hair colour as themselves.
A more extreme example of the breakdown of the $1/N^2$ part
of the likelihood would be if Alice met Bob in Wales and wanted
to infer $f_{\rm Wales}$. Since Bob is probably also Welsh,
the probability of them both being selected would be much greater
than $1/N^2$, and the result would not simplify to
$f_{\rm Wales}^2$.

\section{A Hard Guarantee for Bayesian Inference}\label{sec:guarantee}
In the joint space of parameters and data,
and starting with a joint
prior distribution $p(\theta, D)=p(\theta)p(D \given \theta)$,
conditioning on any true statement $(\theta, D) \in X$ must raise the
amount of probability at the
true value of the pair $(\theta,D)$.
This occurs because the joint posterior is proportional to
\begin{align}
p(\theta, D \given X) &= \frac{p(\theta, D)\mathds{1}\left((\theta, D) \in X\right)}{P(\left(\theta, D) \in X\right)}
\end{align}
where the denominator is less than or equal to 1.
In other words, there is a hard guarantee that conditioning
on any true statement will increase the probability assigned
to the true value of the parameters and the data.
This is in addition to the soft assurance, discussed
in Section~\ref{sec:edge_cases},
that most possible datasets $D$
are not misleading about parameters $\theta$.

\section{Abiogenesis Likelihood Derivation}\label{sec:abiogenesis_derivation}
\begin{figure}
\begin{center}
\includegraphics[width=0.48\textwidth]{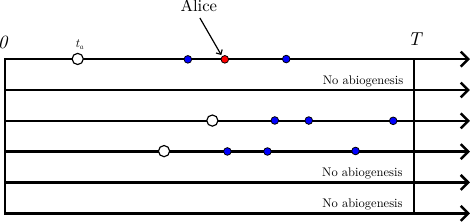}
\end{center}
\caption{The setup for the abiogenesis model. Each habitable planet
has an opportunity for abiogenesis (open circles) and subsequent
observers (filled circles).
\label{fig:abiogenesis_model}
}
\end{figure}

Suppose there are $N$ habitable planets in the universe, each
with a finite time window $[0, T]$ for abiogenesis.
The time of abiogenesis on each planet, $t_{a, i}$, has
an exponential distribution with rate $\lambda$.
If $t_{a, i} \geq T$ for planet $i$,
that just means there isn't any life on that planet.
If $t_{a, i} < T$ so life does arise on planet $i$,
then $m$ observers (reference class members) arise
between time $t_{a, i}$ and $T$.
The setup for this model is depicted in
Figure~\ref{fig:abiogenesis_model}.

In this derivation we make several simplifying assumptions,
for example, we assume the population is $m$ on every life-bearing planet.
It is possible
that the final result may be derivable exactly or approximately under more
general assumptions, but we do not address that possibility here.

The probability that life arises at all on a particular planet is
the probability that an exponential distribution gives a value below $T$:
\begin{align}
1 - \exp(-\lambda T).
\end{align}
The number of planets with life is
\begin{align}
N_{\rm life} &\sim \textnormal{Binomial}(N, 1 - \exp(-\lambda T)).
\end{align}

Alice is defined as being on a planet with abiogenesis time $t_a$ and
existing at $t_{\rm now}$, and having her specific memories. 
Let the population of observers (reference class members)
on all planets with life be $m$, so the total population size is
$M = N_{\rm life}m$.
On each planet with life, we assume the time of occurrence of observers is
given independently by a density $p_t(t \given t_{a, i})$. The
subscript $t$ indicates that it's the probability density for
observer times.

For Alice to exist on planet $i$, the abiogenesis time would have to be $t_a$,
there would have to be an observer at $t_{\rm now}$, and that observer
would have to have Alice's other properties with probability $\epsilon$.
The probability\footnote{Some of the factors here are probability densities
rather than probabilities, but we omit differentials such as $dt_a$
for simplicity as they
do not affect the derivation.}
of all this on planet $i$ is
\begin{align}
P({\rm AE}_i) &= \lambda \exp(-\lambda t_a) m p_t(t_{\rm now} \given t_a)\epsilon.
\end{align}
The probability of it on any planet is therefore
\begin{align}
P({\rm AE}) &=  N\lambda \exp(-\lambda t_a) m p_t(t_{\rm now} \given t_a)\epsilon.\label{eqn:ae}
\end{align}
This would be the likelihood function under FNC and would support large
values of $\lambda$.

We also need the probability for Alice to be selected given that she exists.
If we are allowed to condition on the total population of the universe, $M = mN_{\rm life}$,
this would be simply $1/M$:
\begin{align}
P({\rm AS} \given M, {\rm AE}) &= \frac{1}{M}.
\end{align}
We do not have to handle the $M=0$ case here
because we are conditioning on AE already.
For the FIC likelihood, we need the marginal probability, not conditioning
on $M$:
\begin{align}
P({\rm AS} \given {\rm AE})
    &= \sum_{M=0}^\infty p(M, {\rm AS} \given {\rm AE}) \\
    &= \sum_{M=0}^\infty p(M \given {\rm AE})P({\rm AS} \given M, {\rm AE}) \\
    &= \sum_{M=0}^\infty p(M \given {\rm AE})\frac{1}{M}.\label{eqn:sum}
\end{align}
Note that $p(M \given {\rm AE})$ would only have nonzero probability
at $M=m$, $M=2m$, ..., $Nm$, by the assumption of constant population size.

We can rewrite Equation~\ref{eqn:sum} as a sum over $N_{\rm life}$
rather than $M$:
\begin{align}
P({\rm AS} \given {\rm AE})
    &= \sum_{N_{\rm life}=0}^N p(N_{\rm life} \given {\rm AE})\frac{1}{m N_{\rm life}}.
\end{align}
Applying Bayes' rule to the first factor gives
\begin{align}
P({\rm AS} \given {\rm AE})
    &= \sum_{N_{\rm life}=0}^N \frac{p(N_{\rm life})P({\rm AE} \given N_{\rm life})}{P({\rm AE})}\frac{1}{m N_{\rm life}}.
\end{align}
We already have $P({\rm AE})$ (Equation~\ref{eqn:ae})
which we can substitute into the denominator
of the Bayes' rule expansion:
\begin{align}
P({\rm AS} \given {\rm AE})
    &= \sum_{N_{\rm life}=0}^N \frac{p(N_{\rm life})P({\rm AE} \given N_{\rm life})}{N\lambda \exp(-\lambda t_a) m p_t(t_{\rm now} \given t_a)\epsilon}\frac{1}{m N_{\rm life}}.
\end{align}
The first term in the sum must be zero because $P({\rm AE} \given N_{\rm life})$
must be zero. The sum can start from 1, and many factors not involving $N_{\rm life}$
can be brought outside the sum:
\begin{align}
P({\rm AS} \given {\rm AE})
    &= \frac{\sum_{N_{\rm life}=1}^N \frac{p(N_{\rm life})P({\rm AE} \given N_{\rm life})}{N_{\rm life}}}{Nm\lambda \exp(-\lambda t_a) m p_t(t_{\rm now} \given t_a)\epsilon}.\label{eqn:hard_expectation}
\end{align}

The final difficult term is $P({\rm AE} \given N_{\rm life})$.
This is the probability that Alice exists, if we know the number of life
bearing planets. We can write it down in a similar way to Equation~\ref{eqn:ae}.
If we know there are $N_{\rm life}$ life-bearing planets, the probability
distribution for their abiogenesis times is a truncated exponential of the
form
\begin{align}
\frac{\lambda \exp(-\lambda t_{a, i})}{1 - \exp(-\lambda T)}.
\end{align}
For one of them to match $t_a$ and Alice to exist there, the probability is
\begin{align}
N_{\rm life}\frac{\lambda \exp(-\lambda t_a)}{1 - \exp(-\lambda T)}
m p_t(t_{\rm now} \given t_a)\epsilon.
\end{align}
Substituting this into Equation~\ref{eqn:hard_expectation}
gives
\begin{align}
P({\rm AS} \given {\rm AE})
    &= \frac{\sum_{N_{\rm life}=1}^N \frac{p(N_{\rm life})N_{\rm life}\frac{\lambda \exp(-\lambda t_a)}{1 - \exp(-\lambda T)}
m p_t(t_{\rm now} \given t_a)\epsilon}{N_{\rm life}}}{Nm\lambda \exp(-\lambda t_a) m p_t(t_{\rm now} \given t_a)\epsilon}
     \\
    &= \frac{\frac{\lambda \exp(-\lambda t_a)}{1 - \exp(-\lambda T)}\sum_{N_{\rm life}=1}^N \frac{p(N_{\rm life})N_{\rm life}
m p_t(t_{\rm now} \given t_a)}{N_{\rm life}}}{Nm\lambda \exp(-\lambda t_a) m p_t(t_{\rm now} \given t_a)}
     \\
    &= \frac{m p_t(t_{\rm now} \given t_a)\frac{\lambda \exp(-\lambda t_a)}{1 - \exp(-\lambda T)}\sum_{N_{\rm life}=1}^N p(N_{\rm life})}{Nm\lambda \exp(-\lambda t_a) m p_t(t_{\rm now} \given t_a)}.
\end{align}
The sum is not 1 as it does not include the term $P(N_{\rm life} = 0)$.
The sum is the probability that life exists anywhere at all, which is one
minus the probability that life fails everywhere:
\begin{align}
\sum_{N_{\rm life}=1}^N p(N_{\rm life})
    &= 1 - \exp(-N \lambda T).
\end{align}

Therefore
\begin{align}
P({\rm AS} \given {\rm AE})
    &= \frac{1 - \exp(-N \lambda T)}{Nm\left[1 - \exp(-\lambda T)\right]}.
\end{align}
Multiplying by Equation~\ref{eqn:ae} gives the FIC likelihood:
\begin{align}
P({\rm AE}, {\rm AS}) &=  \left(1 - \exp(-N \lambda T)\right)\frac{\lambda \exp(-\lambda t_a) p_t(t_{\rm now} \given t_a)}{1 - \exp(-\lambda T)}\epsilon.
\end{align}
The $p_t$ term in the numerator is assumed not to depend on $\lambda$ and is
therefore just a constant factor.


\bsp	
\label{lastpage}
\end{document}